# The Vacuum Polarization tensor in 1+1 dimensional space-time.

Dan Solomon
dsolom2@uic.edu
*Department of Physics, University of Illinois at Chicago, 845 West Taylor Street, Chicago, IL 60607.*
December 3, 2017

## Abstract.

Quantum field theory (QFT) is supposed to be gauge invariant. However it has been well established that a direct calculation of the vacuum polarization tensor produces a non-gauge invariant result. In this paper it will be shown that this problem is due to the fact that there is a lower bound to the free field energy in QFT. The vacuum polarization tensor will be calculated in 1+1 dimensional space-time and is shown not to be gauge invariant. The gauge invariance of the theory can be restored through a regularization procedure which eliminates the non-gauge invariant terms. However it will be shown that this will impact the free field energy. If the free field energy is defined so that the vacuum state has an energy of zero then the impact of regularization is to introduce states whose free field energy is less than zero.

## 1. Introduction.

It has been well established that quantum field theory (QFT) contains anomalies [1,2,3]. These occur when the result of some calculation does not agree with an underlying classical symmetry of the theory. This is the case with gauge invariance [4,5,6]. A change in the gauge is a change in the electric potential that does not result in a change in the electromagnetic field. The electromagnetic field, $\vec{E}$ and $\vec{B}$, is given in terms of the electric potential, $\left(A_0, \vec{A}\right)$ by,

$$\vec{E} = -\left(\frac{\partial \vec{A}}{\partial t} + \vec{\nabla} A_0\right), \quad \vec{B} = \vec{\nabla} \times \vec{A} \tag{1.1}$$

A change in the electric potential that does not result in a change in the electromagnetic field is given by,

$$\vec{A} \to \vec{A}' = \vec{A} - \vec{\nabla}\chi, \quad A_0 \to A_0' = A_0 + \frac{\partial \chi}{\partial t} \tag{1.2}$$

In the above $\chi(x)$ is an arbitrary real valued function.

Physical theories are assumed to be gauge invariant [7,8]. In order for QFT to be gauge invariant a change in the gauge should not produce a change in some physical observable such as the charge and current density, for example. However it is well know that when the vacuum current is calculated the results include non-gauge invariant terms. (See discussion in [4]).

The first order change in the vacuum current, due to the application of an electromagnetic field, is given by,

$$J^{\mu}_{vac}(x) = \int \pi^{\mu\nu}(x - x')A_{\nu}(x')d^{4}x' \tag{1.3}$$

where $\pi^{\mu\nu}$ is the vacuum polarization tensor and summation over repeated indices is implied. The above relationship can be expressed in terms of Fourier transformed functions as,

$$J^{\mu}_{vac}(k) = \pi^{\mu\nu}(k)A_{\nu}(k) \tag{1.4}$$

For the Fourier transformed electric potential the gauge transformation takes the form,

$$A_{\nu}(k) \to A'_{\nu}(k) = A_{\nu}(k) + ik_{\nu}\chi(k) \tag{1.5}$$

The change in the vacuum current $\delta_{g}J^{\mu}_{vac}(k)$ due to a gauge transformation is then given by,

$$\delta_{g}J^{\mu}_{vac}(k) = ik_{\nu}\pi^{\mu\nu}(k)\chi(k) \tag{1.6}$$

Therefore, for the theory to be gauge invariant, $\delta_{g}J^{\mu}_{vac}(k)$ must be zero which means that the vacuum polarization tensor must satisfy,

$$k_{\nu}\pi^{\mu\nu}(k) = 0 \tag{1.7}$$

However, when the vacuum polarization tensor is calculated it is found that the above relationship does not hold. An example of this is a calculation by Heitler [9] which shows that,

$$\pi^{\mu\nu}(k) = \pi^{\mu\nu}_{G}(k) + \pi^{\mu\nu}_{NG}(k) \tag{1.8}$$

where,

$$\pi^{\mu\nu}_{G}(k) = -\left(\frac{2e^{2}}{3\pi}\right)\left(k^{\mu}k^{\nu} - g^{\mu\nu}k^{2}\right) \times \int_{2m}^{\infty} dz \frac{\left(z^{2} + 2m^{2}\right)\sqrt{z^{2} - 4m^{2}}}{z^{2}\left(z^{2} - k^{2}\right)} \tag{1.9}$$

and,

$$\pi^{\mu\nu}_{NG}(k) = -\left(\frac{2e^{2}}{3\pi}\right)g^{\mu}_{\nu}\left(1 - g^{\mu 0}\right) \times \int_{2m}^{\infty} dz \frac{\left(z^{2} + 2m^{2}\right)\sqrt{z^{2} - 4m^{2}}}{z^{2}} \tag{1.10}$$

where there is no summation over the two $\mu$ superscripts that appear on the right in this last equation. It can be readily shown that $k_{\nu}\pi^{\mu\nu}_{G}(k) = 0$ but that $k_{\nu}\pi^{\mu\nu}_{NG}(k) \neq 0$. Therefore $\pi^{\mu\nu}(k)$ can be separated into two parts. One part, $\pi^{\mu\nu}_{G}$, is gauge invariant and the other part, $\pi^{\mu\nu}_{NG}$, is not gauge invariant. In order to obtain a gauge invariant theory the non-gauge invariant term must be removed.

Another example of this problem is from Section 14.2 of Greiner et al [8]. For the vacuum polarization tensor they obtain,

$$\pi^{\mu\nu}(k)=\left(g^{\mu\nu}k^2-k^{\mu}k^{\nu}\right)\pi\left(k^2\right)+g^{\mu\nu}\pi_{sp}\left(k^2\right) \tag{1.11}$$

where $\pi\left(k^2\right)$ and $\pi_{sp}\left(k^2\right)$ are given in [8]. In order for the above expression to be gauge invariant $\pi_{sp}\left(k^2\right)$ must equal to zero however it is shown in [8] that this is not the case. Many other examples that demonstrate the existence of this problem are available from the literature including Section 8 of W. Pauli [10] and [11,12,13]. (Also see the discussion in Ref. [4]).

In order to obtain a gauge invariant result the non-gauge invariant terms must be removed. This can simply be done by "hand", that is the offending terms are simply removed (as is the case with Heitler [9] and Greiner et al [8]) or more sophisticated mathematical techniques can be employed to produce a gauge invariant result. These techniques are generally referred to as "regularization". One such technique is Pauli-Villars regularization. An example of this is given in a calculation by Griener and Reinhardt [14]. In this case an "auxiliary" fermion field with infinite mass is introduced. The vacuum polarization tensor of this auxiliary field is then subtracted from the final result to obtain the "regularized" vacuum polarization tensor. That is, suppose we are calculating the polarization tensor of a fermion field of mass $m$. We initially obtain $\pi^{\mu\nu}(k;m)$ which will not be gauge invariant. The regularized gauge invariant result is obtained from the expression,

$$\pi^{\mu\nu}_{REG}(k;m)=\pi^{\mu\nu}(k;m)-\pi^{\mu\nu}(k;M) \tag{1.12}$$

where $M\to\infty$ is the mass of the auxiliary field.

Another such technique is dimensional regularization. In dimensional regularization quantities are calculated in $4-\varepsilon$ dimensions, instead of $4$, where $\varepsilon\to 0$. This will produce a gauge invariant result for the vacuum polarization tensor. A detailed calculation of the vacuum polarization tensor using this method is given in [15].

In their discussion of this problem Greiner et al write (page 398 of [8]) that since the vacuum current "is, in principle, observable, this latter term [i.e. $\pi_{sp}\left(k^2\right)$ in Eq. (1.11)] violates the gauge invariance of the theory. This would be a severe contradiction to the experimentally confirmed gauge independence of QED. Thus in a satisfactory formulation of the theory of charge particles, such a term ( $\pi_{sp}\left(k^2\right)$ ) must not appear. We now explicitly show that it is indeed present showing that QED is not a complete theory. As one counter example or inconsistency suffices to prove a theory wrong, we should, in principle, spend the rest of this book searching for an improved theory. However, there is little active work on this today because (1) there is a common belief that some artifact of the exact mathematics is the source of this problem; (2) this problem may disappear when a properly generalized theory, including in its framework all charge Dirac particles, is achieved."

Although "regularization" seems to cure this problem there is no good explanation, in my opinion, on why this problem occurs in the first place. Why do these non-gauge invariant terms appear in a theory that is supposed to be gauge invariant? Why don't calculations of the vacuum polarization tensor simply produce a gauge invariant result in the first place? The main purpose of this paper is to show that the common belief suggested by Greiner that "some artefact of the exact mathematics is a source of the problem" is not true. I will demonstrate that the calculations that result in a non-gauge invariant theory are, indeed, correct and are not an artifact of the mathematics. In fact one can show that the theory is not gauge invariant based on it's theoretical formulation. This has been demonstrated previously in Ref. [4,5]. Therefore that fact that non-gauge invariant terms appear should not be a surprise but should be an expected result based on the underlying formulation of the theory.

In addition it will be shown that when the theory is regularized to remove the gauge invariant terms then other quantities to the theory may be impacted. In particular we will look at the energy in the absence of external potentials which is commonly called the free field energy. Normally the theory can be set up so that the vacuum state $|0\rangle$ has a free field energy of zero. However, the use of regularization to remove the non-gauge invariant terms, will result in the existence of quantum states with a free energy that is less than zero. So one result of regularization is that the vacuum state is no longer the lower bound to the free field energy.

In this paper the previous results of [4,5] will be reviewed and re-derived. Next, the vacuum polarization tensor with be calculated in 1+1 dimensional space-time. The reason for working in 1+1 dimensions is that no divergences occur and all calculations are well-defined. It will be shown that the vacuum polarization tensor is given by $\pi_{\mu\nu} = \pi_{\mu\nu,G} + \pi_{\mu\nu,NG}$ where $\pi_{\mu\nu,G}$ is gauge invariant and $\pi_{\mu\nu,NG}$ is not gauge invariant. The purpose of this calculation is to demonstrate that the existence of this non-gauge invariant term is not an "artifact of the exact mathematics" but is the correct result of a straightforward mathematical calculation.

In order to achieve a gauge invariant theory the non-gauge invariant part $\pi_{\mu\nu,NG}$ must be removed from the result. This can be done be simply redefining the vacuum polarization tensor as $\bar{\pi}_{\mu\nu} = \pi_{\mu\nu,G}$. Next, the impact of this change on the free field energy will be examined. In the normal formulation of QFT the vacuum state vector $|0\rangle$ is the lower bound to the free field energy. However it will be shown that the act of regularizing the vacuum polarization tensor may impact other calculations and this includes the free field energy. It will be shown that, as a result of regularization, the vacuum state is no longer a lower bound to the free field energy. Finally we will examine how the use of Pauli-Villars regularization impacts this problem.

## 2. Four Elements of QFT.

In this section we will consider a simple "toy model" field theory consisting of non-interacting fermions in the presence of a non-quantized classical electromagnetic field. Four basic elements that are normally considered to be part of QFT will be presented. We will consider them to be postulates in that they will be presented as being true without proof. In the next section it will be shown that these elements are mathematically inconsistent. In later sections it will be shown that this inconsistency leads to the anomaly involving gauge invariance. With some modifications the following discussion will rely heavily on material from Refs [4] and [5].

We will start by working in the Schrödinger representation. Natural units are used so that $\hbar = c = 1$. The first element of the theory is that the time evolution of the state vector $|\Omega(t)\rangle$ and its dual $\langle\Omega(t)|$ are given by the Schrödinger equation,

$$\frac{\partial|\Omega(t)\rangle}{\partial t} = -i\hat{H}(t)|\Omega(t)\rangle, \quad \frac{\partial\langle\Omega(t)|}{\partial t} = i\langle\Omega(t)|\hat{H}(t). \tag{2.1}$$

$\hat{H}(t)$ is the Hamiltonian operator which is given by,

$$\hat{H}(t) = \hat{H}_0 - \int \hat{\vec{J}}_S(\vec{x}) \cdot \vec{A}(\vec{x},t)d\vec{x} + \int \hat{\rho}_S(\vec{x})A_0(\vec{x},t)d\vec{x}. \tag{2.2}$$

$\hat{\vec{J}}_S(\vec{x})$ and $\hat{\rho}_S(\vec{x})$ are time independent Schrodinger operators for the current and charge density, respectively, and the electric potential $(A_0, \vec{A})$ are unquantized classical real valued functions. The state vector $|\Omega(t)\rangle$ is assumed to be normalized so that $\langle\Omega(t)|\Omega(t)\rangle = 1$. $\hat{H}_0$ is the free field Hamiltonian operator which is the Hamiltonian when the electric potentials are zero.

In addition to the above a second element of the theory is the assumption of gauge invariance. As a result of this assumption a change in the gauge will not result in a change to the current and charge expectation values. These are defined by,

$$\vec{J}_e(\vec{x},t) = \langle\Omega(t)|\hat{\vec{J}}_S(\vec{x})|\Omega(t)\rangle, \quad \rho_e(\vec{x},t) = \langle\Omega(t)|\hat{\rho}_S(\vec{x})|\Omega(t)\rangle. \tag{2.3}$$

A third element, that is a normal part of QFT, is local conservation of electric charge which is the continuity equation,

$$\frac{\partial \rho_e(\vec{x},t)}{\partial t} + \vec{\nabla} \cdot \vec{J}_e(\vec{x},t) = 0 \tag{2.4}$$

A fourth element of the theory is that there exists a minimum value to the free field energy. The free field energy $\xi_f(|\Omega\rangle)$ of the normalized state vector $|\Omega\rangle$ is defined to be,

$$\xi_f(|\Omega\rangle) = \langle\Omega|\hat{H}_0|\Omega\rangle \tag{2.5}$$

Let $|n\rangle$ be the set of orthonormal eigenstates of $\hat{H}_0$ with eigenvalues $\varepsilon_n$ so that,

$$\hat{H}_0|n\rangle = \varepsilon_n|n\rangle \tag{2.6}$$

The eigenvector with the smallest eigenvalue is called the vacuum state, $|0\rangle$, with eigenvalue value $\varepsilon_0 = 0$. Therefore,

$$\varepsilon_n > \varepsilon_0 = 0 \text{ for } |n\rangle \neq |0\rangle \tag{2.7}$$

Since any arbitrary normalized state $|\Omega\rangle$ can be expanded as a sum of eigenstates $|n\rangle$ it can be shown that,

$$\xi_f(|\Omega\rangle) \geq \xi_f(|0\rangle) = 0 \text{ for all } |\Omega\rangle \tag{2.8}$$

To sum up we have introduced four elements of QFT. They are: (1) the Schrodinger equation, (2) the principle of gauge invariance, (3) the continuity equation, and (4) there is a minimum to the free field energy. It will be shown in the next section that these four elements are not mathematically consistent.

## 3. A Mathematical Inconsistency.

Recall that the free field energy of the state vector $|\Omega(t)\rangle$ is given by $\xi_f(t) = \langle\Omega(t)|\hat{H}_0|\Omega(t)\rangle$. Using (2.2) this can be expressed as,

$$\xi_f(t) = \langle\Omega(t)|\left\{\hat{H}(t) - \left(-\int \hat{\vec{J}}_S(\vec{x}) \cdot \vec{A}(\vec{x},t)d\vec{x} + \int \hat{\rho}_S(\vec{x}) A_0(\vec{x},t)d\vec{x}\right)\right\}|\Omega(t)\rangle \tag{3.1}$$

Using (2.3) in the above this can be written as,

$$\xi_f(t) = \langle\Omega(t)|\hat{H}(t)|\Omega(t)\rangle - \left(-\int \vec{J}_e(\vec{x},t) \cdot \vec{A}(\vec{x},t)d\vec{x} + \int \hat{\rho}_e(\vec{x},t) A_0(\vec{x},t)d\vec{x}\right) \tag{3.2}$$

It is shown in [5] that we can use (2.1) in the above to obtain,

$$\frac{d\xi_f(t)}{dt} = \int \frac{\partial \vec{J}_e(\vec{x},t)}{\partial t} \cdot \vec{A}(\vec{x},t)d\vec{x} - \int \frac{\partial \hat{\rho}_e(\vec{x},t)}{\partial t} A_0(\vec{x},t)d\vec{x} \tag{3.3}$$

Next assume that,

$$\vec{A}(\vec{x},t) = -\nabla\chi(\vec{x},t), \quad A_0(\vec{x},t) = \partial\chi(\vec{x},t)/\partial t \tag{3.4}$$

where $\chi(\vec{x},t)$ is a real valued function subject to the initial condition at some initial time $t_a$,

$$\chi(\vec{x},t_a) = 0, \quad \partial\chi(\vec{x},t_a)/\partial t = 0 \tag{3.5}$$

Use this in (3.3) to obtain,

$$\frac{d\xi_f(t)}{dt} = -\int \frac{\partial \vec{J}_e(\vec{x},t)}{\partial t} \cdot \nabla\chi(\vec{x},t) d\vec{x} - \int \frac{\partial \hat{\rho}_e(\vec{x},t)}{\partial t} \frac{\partial \chi(\vec{x},t)}{\partial t} d\vec{x} \tag{3.6}$$

Assume reasonable boundary conditions and integrate by parts to obtain,

$$\frac{d\xi_f(t)}{dt} = \int \chi(\vec{x},t) \frac{\partial \nabla \cdot \vec{J}_e(\vec{x},t)}{\partial t} d\vec{x} - \int \frac{\partial \hat{\rho}_e(\vec{x},t)}{\partial t} \frac{\partial \chi(\vec{x},t)}{\partial t} d\vec{x} \tag{3.7}$$

This can be rewritten as,

$$\frac{d\xi_f(t)}{dt} = \int \chi(\vec{x},t) \frac{\partial \nabla \cdot \vec{J}_e(\vec{x},t)}{\partial t} d\vec{x} - \frac{\partial}{\partial t}\left( \int \frac{\partial \hat{\rho}_e(\vec{x},t)}{\partial t} \chi(\vec{x},t) d\vec{x} \right) + \int \frac{\partial^2 \hat{\rho}_e(\vec{x},t)}{\partial t^2} \chi(\vec{x},t) d\vec{x} \tag{3.8}$$

Rearrange terms to obtain,

$$\frac{d\xi_f(t)}{dt} = \int \chi(\vec{x},t) \frac{\partial}{\partial t}\left( \frac{\partial \hat{\rho}_e(\vec{x},t)}{\partial t} + \nabla \cdot \vec{J}_e(\vec{x},t) \right) d\vec{x} - \frac{\partial}{\partial t}\left( \int \frac{\partial \hat{\rho}_e(\vec{x},t)}{\partial t} \chi(\vec{x},t) d\vec{x} \right) \tag{3.9}$$

From the discussion in the last section it was noted that one of the elements of QFT is that the continuity equation holds. Therefore use Eq. (2.4) in the above to obtain,

$$\frac{d\xi_f(t)}{dt} = -\frac{\partial}{\partial t}\left( \int \frac{\partial \hat{\rho}_e(\vec{x},t)}{\partial t} \chi(\vec{x},t) d\vec{x} \right) \tag{3.10}$$

Once again use (2.4) to obtain,

$$\frac{d\xi_f(t)}{dt} = \frac{\partial}{\partial t}\left( \int \chi(\vec{x},t) \nabla \cdot \vec{J}_e(\vec{x},t) d\vec{x} \right) \tag{3.11}$$

Next integrate with respect to time from the initial time $t_a$ to some final time $t_b$ and use the initial condition (3.5) to obtain,

$$\xi_f(t_b) = \xi_f(t_a) + \left( \int \chi(\vec{x},t_b) \nabla \cdot \vec{J}_e(\vec{x},t_b) d\vec{x} \right) \tag{3.12}$$

If we use (3.4) in (1.1) it is evident that the electromagnetic field is zero ($\vec{E} = 0$ and $\vec{B} = 0$) so that the electromagnetic field is independent of $\chi(\vec{x},t)$. Therefore if we use the assumption that the theory is gauge invariant we conclude that $\vec{J}_e(\vec{x},t)$ must also be independent of $\chi(\vec{x},t)$. Next, assume that we have set up the state vector so that $\nabla\cdot\vec{J}_e(\vec{x},t_b)\neq 0$. We are free to specify $\chi(\vec{x},t_b)$ so let it be given by $\chi(\vec{x},t_b)=-f\nabla\cdot\vec{J}_e(\vec{x},t_b)$ where $f$ is a positive constant. Use this in (3.12) to obtain,

$$\xi_f(t_b)=\xi_f(t_a)-f\int\left(\nabla\cdot\vec{J}_e(\vec{x},t_b)\right)^2 d\vec{x} \tag{3.13}$$

From this result it is evident that there is no lower bound to $\xi_f(t_b)$ since the integral is positive and we can always let $f\rightarrow+\infty$. We have just shown that for a gauge invariant theory there is no lower bound to the free field energy $\xi_f$. However this contradicts Eq. (2.8) which states that the vacuum state $|0\rangle$ is a lower bound to the free field energy.

Therefore there is a mathematical inconsistency between the requirement that there must be a lower bound to the free field energy and the requirement that theory be gauge invariant. If we then, quantize the theory in the usual way such that there is a lower bound to the free field energy (that is Eq. (2.8) is true) then the theory will no longer be gauge invariant. This is, then, why all these calculations of the vacuum polarization tensor that have been discussed above yield non-gauge invariant results which have to be corrected by the additional step of regularization. Therefore this problem is not an "artifact" of the mathematics but is built into the theory at the fundamental level. This will be further demonstrated below where the vacuum polarization tensor will be calculated for our simplified "toy model" in 1+1 dimensional space-time and shown to contain non-gauge invariant terms.

## 4. Defining the field operator.

In this section we will introduce some additional elements that are necessary to the formulation of out "toy model" field theory. This includes the definition of field operators as an expansion of creation and annihilation operators. This will allow us to calculate the vacuum polarization tensor in order to determine if it is gauge invariant. In the rest of this paper we will work in 1+1 dimensional space-time. The advantage of working in 1+1 dimensions is all the quantities we calculate in the following will be finite and well defined. When we work in higher dimensions divergences occur which complicate the discussion. In the following $\sigma_i$ are the Pauli matrices, $x$ will be the space dimension, and $t$ will be the time dimension.

It is convenient to rewrite Eq. (2.2) as,

$$\hat{H}(t)=\hat{H}_0+\hat{V}(t),\quad \hat{V}(t)=\int J_{S,\nu}(x)\cdot A^{\nu}(x,t)dx, \tag{4.1}$$

where $\hat{J}_{S,\nu}A^{\nu}=\hat{J}_{S,0}A_0-\hat{J}_{S,1}A_1$ with $\hat{J}_{S,1}(x)$ and $\hat{J}_{S,0}(x)$ the current and charge operators, respectively, in the Schrodinger picture. The time-independent Schrodinger field operators are given by,

$$\hat{\psi}_S(x)=\sum_p\left(\hat{b}_p u_p e^{ipx}+\hat{d}_p^{\dagger}v_p e^{-ipx}\right),\quad \hat{\psi}_S^{\dagger}(x)=\sum_p\left(\hat{b}_p^{\dagger}u_p^{\dagger}e^{-ipx}+\hat{d}_p v_p^{\dagger}e^{+ipx}\right). \tag{4.2}$$

The summation is over the momentum $p=2\pi n/L$ where $n$ is an integer and where $L\to\infty$ is the one-dimensional integration volume. Also $\hat{b}_p$ and $\hat{b}_p^{\dagger}$ are the electron destruction and creation operators, respectively, whereas $\hat{d}_p$ and $\hat{d}_p^{\dagger}$ play the same roles for positrons. These operators obey the usual anti-commutation relationships, $\left\{\hat{b}_p,\hat{b}_q^{\dagger}\right\}=\left\{\hat{d}_p,\hat{d}_q^{\dagger}\right\}=\delta_{pq}$ with all other anti-commutators being equal to zero. The 2-spinors $u_p$ and $v_p$ are solutions to,

$$H_0\left(u_p e^{ipx}\right)=E_p u_p e^{ipx},\qquad H_0\left(v_p e^{-ipx}\right)=-E_p v_p e^{-ipx} \tag{4.3}$$

where $E_p=\sqrt{p^2+m^2}$ and $H_0=\left(-i\sigma_1\frac{\partial}{\partial x}+m\sigma_3\right)$. This yields,

$$u_p=N_p\begin{pmatrix}1\\ p/\left(E_p+m\right)\end{pmatrix},\ v_p=N_p\begin{pmatrix}p/\left(E_p+m\right)\\ 1\end{pmatrix},\ N_p=\sqrt{\frac{\left(E_p+m\right)}{2LE_p}} \tag{4.4}$$

The operators $\hat{H}_0$, $\hat{J}_{S,1}(x)$, and $\hat{J}_{S,0}(x)$ are given in terms of the field operators by,

$$\hat{H}_0=\int dx:\hat{\psi}_S^{\dagger}(x)H_0\hat{\psi}_S(x):,\quad \hat{J}_{S,1}(x)=e:\hat{\psi}_S^{\dagger}(x)\sigma_1\hat{\psi}_S(x):,\quad \hat{J}_{S,0}(x)=e:\hat{\psi}_S^{\dagger}(x)\hat{\psi}_S(x): \tag{4.5}$$

where the colons indicate normal order and $e$ is the charge on the electron. When these relationships are used we obtain,

$$\hat{H}_0=\sum_p E_p\left(\hat{b}_p^{\dagger}\hat{b}_p+\hat{d}_p^{\dagger}\hat{d}_p\right) \tag{4.6}$$

The vacuum state $|0\rangle$ is the state which is destroyed by the destruction operators, $\hat{b}_p|0\rangle=\hat{d}_p|0\rangle=0$. New states are formed by acting on the vacuum state $|0\rangle$ with some combination of creation operators $\hat{b}_p^{\dagger}$ and $\hat{d}_p^{\dagger}$. The result is that for any arbitrary normalized state vector $|\Omega\rangle$ the free field energy $\langle\Omega|\hat{H}_0|\Omega\rangle\geq 0$. However, as was demonstrated in Section 3, the theory cannot be both gauge invariant and have a lower bound to the free field energy. But it has just been shown that the free field energy has a lower bound of zero. Therefore we would expect that the theory should not be gauge invariant. This will be demonstrated in the next section where we will calculate the vacuum polarization tensor and show that it is not gauge invariant.

## 5. Vacuum Polarization tensor.

In this section the vacuum polarization tensor will be calculated for our 1+1 dimensional problem. Assume at $t \to -\infty$ the system is in the vacuum state $|0\rangle$ and the electric potential is zero. Next apply a non-zero electric potential. The state vector will evolve in time per the Schrodinger equation. If we assume that the electric potential is small then the state vector at some future time can be determined as a perturbative expansion. By the usual methods of perturbation theory we obtain [12],

$$|\Omega_I(t)\rangle = \left(1 - i\int_{-\infty}^{t} \hat{V}_I(t')dt' - \int_{-\infty}^{t} \hat{V}_I(t')dt' \int_{-\infty}^{t'} \hat{V}_I(t'')dt'' + \ldots\right)|0\rangle \tag{5.1}$$

where $|\Omega_I(t)\rangle$ is the state vector in the interaction picture and $\hat{V}_I(t) = e^{+i\hat{H}_0 t}\hat{V}(t)e^{-i\hat{H}_0 t}$. The interaction current operator is given by $\hat{J}_\nu(x,t) = e^{+i\hat{H}_0 t}\hat{J}_{S,\nu}(x)e^{-i\hat{H}_0 t}$. The vacuum current is the current that is induced in the vacuum by the application of the electric potential and is given by,

$$J_{\mu,vac}(x,t) = \langle\Omega_I(t)|\hat{J}_\mu(x,t)|\Omega_I(t)\rangle \tag{5.2}$$

Next use (5.1) in the above to show that to the lowest order in the expansion the vacuum current is,

$$J_{\mu,vac}(x,t) = -i\int dx_1 \int_{-\infty}^{t} dt_1 \langle 0|\left[\hat{J}_\mu(x,t), \hat{J}_\nu(x_1,t_1)\right]|0\rangle A^\nu(x_1,t_1) \tag{5.3}$$

This can be written as,

$$J_{\mu,vac}(x,t) = \int dx_1 \int_{-\infty}^{+\infty} dt_1 \pi_{\mu\nu}(x - x_1, t - t_1) A^\nu(x_1,t_1) \tag{5.4}$$

Where the vacuum polarization tensor is defined as,

$$\pi_{\mu\nu}(x - x_1, t - t_1) = -i\langle 0|\left[\hat{J}_\mu(x,t), \hat{J}_\nu(x_1,t_1)\right]|0\rangle \theta(t - t_1) \tag{5.5}$$

There is another definition of the vacuum polarization tensor that is also used in the literature. This definition is,

$$\pi_{T,\mu\nu}(x - x_1, t - t_1) = -i\langle 0|T\left[\hat{J}_\mu(x,t)\hat{J}_\nu(x_1,t_1)\right]|0\rangle \tag{5.6}$$

where $T$ stands for time-ordered product. $\pi_{T,\mu\nu}$ is used in the calculation of the first order correction to the photon propagator. The focus of this paper is on $\pi_{\mu\nu}$ (Eq. (5.5)) however $\pi_{T,\mu\nu}$ will be briefly considered also.

We are interesting in evaluating the vacuum polarization tensor $\pi_{\mu\nu}(x,t)$ in momentum space,

$$\pi_{\mu\nu}(k_1,k_0) = \int \pi_{\mu\nu}(x,t) e^{-ik_0 t} e^{+ik_1 x} dx dt \tag{5.7}$$

Using the information in Section 4 and the above relationships it is shown in Appendix 1 that,

$$\pi_{\mu\nu}(k_1,k_0) = \pi_{\mu\nu,G}(k_1,k_0) + \pi_{\mu\nu,NG}(k_1,k_0) \tag{5.8}$$

where,

$$\pi_{\mu\nu,G}(k_1,k_0)=(k_\mu k_\nu - g_{\mu\nu}k^2)\Pi(k^2),\quad \pi_{\mu\nu,NG}(k_1,k_0)=g^1_\mu g^1_\nu \Gamma \tag{5.9}$$

where $g^1_\rho = 1$ if $\rho = 1$ and $g^1_\rho = 0$ if $\rho = 0$, $k^2 = k_0{}^2 - k_1{}^2$, $g_{\mu\nu} = \begin{pmatrix} 1 & 0 \\ 0 & -1 \end{pmatrix}$ and,

$$\Pi(k^2) = -\frac{e^2m^2}{8\pi}\int_{m^2}^{\infty} ds\left(\frac{1}{s^2\sqrt{1-(m^2/s)}}\right)\left(\frac{1}{s-(k^2/4)+(k_0 i\varepsilon/2)}\right) \tag{5.10}$$

$$\Gamma = \frac{-e^2m^2}{2\pi}\int_{m^2}^{\infty} ds\left(\frac{1}{s^2\sqrt{1-(m^2/s)}}\right) = -\frac{e^2}{\pi} \tag{5.11}$$

where $\varepsilon \to 0$ from above. From Appendix 1 we also show that $\Pi(k^2)$ can be expressed as,

$$\Pi(k^2) = -\frac{e^2}{\pi}\int_0^1\left(\frac{y(1-y)dy}{m^2 - k^2 y(1-y) + k_0 i\varepsilon}\right) \tag{5.12}$$

Recall that in order to be gauge invariant the vacuum polarization tensor must satisfy $k^u \pi_{\mu\nu}(k_1,k_0) = 0$. Note that $k^u \pi_{\mu\nu,G}(k_1,k_0) = 0$ however $k^u \pi_{\mu\nu,NG}(k_1,k_0) = k^1 g^1_\nu \Gamma \neq 0$ if $\nu = 1$. Therefore we have found that the vacuum polarization tensor $\pi_{\mu\nu}(k_1,k_0)$ is not gauge invariant. It consists of two terms where $\pi_{\mu\nu,G}(k_1,k_0)$ is the gauge invariant part and $\pi_{\mu\nu,NG}(k_1,k_0)$ is the non-gauge invariant part. One key result is that this non-gauge invariant term is not due to an artifact of the mathematics but is the result of a straight forward mathematical calculation. Also, based on are previous discussion, we should have expected such a result because, as was shown in Section 3 there is a mathematical inconsistency between the principle of gauge in invariance and the fact that there is a lower bound to the free field energy. However in Section 4 we defined the mathematical structure of the theory in such a way that there is a lower bound to the free field energy which is zero. Therefore we expect that this theory will not be gauge invariant which has just been shown to be the case.

So how do we obtain a gauge invariant theory? One way is to simply drop the non-gauge invariant term from the solution. In this case the regularized vacuum polarization tensor is,

$$\bar{\pi}_{\mu\nu}(k_1,k_0) = \pi_{\mu\nu}(k_1,k_0) - \pi_{\mu\nu,NG}(k_1,k_0) = \pi_{\mu\nu,G}(k_1,k_0) \tag{5.13}$$

$\bar{\pi}_{\mu\nu}(k_1,k_0)$ is now defined to be the vacuum polarization tensor and it is gauge invariant. Converting to coordinate space we obtain,

$$\bar{\pi}_{\mu\nu}(x,t) = \pi_{\mu\nu}(x,t) - \int \frac{dk_0 dk_1}{(2\pi)^2} g_\mu^1 g_\nu^1 \Gamma e^{+ik_0 t} e^{-ik_1 x} = \pi_{\mu\nu}(x,t) - g_\mu^1 g_\nu^1 \Gamma \delta(x)\delta(t) \tag{5.14}$$

Another way to obtain a gauge invariant result is to use Pauli-Villars regularization which is discussed in Section 7. The results for $\pi_{T,\mu\nu}$, as defined in Eq. (5.6), are given in Appendix 4.

## 6. The Free Field Energy.

At this point we have obtained a gauge invariant vacuum polarization tensor. However recall from the discussion in Section 3 that there is an inconsistency between the theory being gauge invariant and the assumption that there is a lower bound to the free field energy. Therefore at this point we examine how achieving a gauge invariant theory impacts on the free field energy.

Consider the state $|\Omega(0)\rangle$ where $|\Omega(t)\rangle = e^{i\hat{H}_0 t}|\Omega_I(t)\rangle$ with $|\Omega_I(t)\rangle$ given by Eq. (5.1). This yields,

$$|\Omega(0)\rangle = \left(1 - i\int_{-\infty}^{0} V_I(t')dt' - \int_{-\infty}^{0} V_I(t')dt' \int_{-\infty}^{t'} V_I(t'')dt'' + \ldots\right)|0\rangle \tag{6.1}$$

The free field energy for the state $|\Omega(0)\rangle$ is given by $\xi_f = \langle\Omega(0)|\hat{H}_0|\Omega(0)\rangle$. Use this along with $\hat{H}_0|0\rangle = \langle 0|\hat{H}_0 = 0$ to obtain for the lowest order term of the free field energy,

$$\xi_f = \langle 0|\left(\int_{-\infty}^{0} \hat{V}_I(t)dt\right)\hat{H}_0\left(\int_{-\infty}^{0} \hat{V}_I(t)dt\right)|0\rangle \tag{6.2}$$

With $\hat{H}_0$ defined per Eq. (4.6) we know that the above quantity must be greater or equal to zero. This will be shown by direct calculation for a specific case. Consider the case where $A_0(x,t) = 0$ and $A_1(x,t) = fe^{\lambda t}/\sqrt{L}$ where $f$ and $\lambda$ are positive constants and $L$ is the one dimensional integration volume. In this case (6.2) becomes,

$$\xi_f = (f^2/L)\langle 0|\left(\int_{-\infty}^{0} e^{\lambda t}dt \int dx \hat{J}_1(x,t)\right)\hat{H}_0\left(\int_{-\infty}^{0} e^{\lambda t}dt \int dx \hat{J}_1(x,t)\right)|0\rangle \tag{6.3}$$

This can be readily evaluated using the definitions in Section 4. It is shown in Appendix 2 that this is equal to,

$$\xi_f = \frac{f^2 e^2 m^2}{4\pi} \int_{-\infty}^{+\infty} \left(\frac{1}{E_p\left(E_p^{\,2} + (\lambda/2)^2\right)}\right) dp \tag{6.4}$$

This is obviously greater than zero. However this calculation did not take into account the effects of regularization. Recall that when we calculated the vacuum polarization tensor $\pi_{\mu\nu}$ we discovered that in order to obtain a gauge invariant theory we had to replace $\pi_{\mu\nu}$ with $\bar{\pi}_{\mu\nu}$. Therefore, for mathematical consistency, it is required that whenever $\pi_{\mu\nu}$ appears in a calculation it should be replaced by $\bar{\pi}_{\mu\nu}$. Now it is not obvious that $\pi_{\mu\nu}$ appears in (6.2) however in the following it will be shown below that it does, indeed, appear.

Using $\hat{H}_0|0\rangle = \langle 0|\hat{H}_0 = 0$ in (6.2) we can write,

$$\xi_f = -\frac{1}{2}\langle 0|\left[\left[\hat{H}_0, \left(\int_{-\infty}^{0}\hat{V}_I(t)dt\right)\right], \left(\int_{-\infty}^{0}\hat{V}_I(t)dt\right)\right]|0\rangle \tag{6.5}$$

Next simplify the problem by setting $A_0(x,t)=0$ to obtain $\hat{V}_I(t) = -\int \hat{J}_1(x,t)A_1(x,t)dx$. Use this in the above expression to obtain,

$$\xi_f = -\frac{1}{2}\int dx_1\int dx_2 \int_{-\infty}^{0} dt_1 \int_{-\infty}^{0} dt_2 A_1(x_1,t_1)A_1(x_2,t_2)\langle 0|\left[\left[\hat{H}_0, \hat{J}_1(x_1,t_1)\right], \hat{J}_1(x_2,t_2)\right]|0\rangle \tag{6.6}$$

Recall that $\hat{J}_1(x,t) = e^{i\hat{H}_0 t}\hat{J}_{S,1}(x)e^{-i\hat{H}_0 t}$ which yields $\partial\hat{J}_1(x,t)/\partial t = i\left[\hat{H}_0, \hat{J}_1(x,t)\right]$. Use this in the above to obtain,

$$\xi_f = \frac{i}{2}\int dx_1\int dx_2 \int_{-\infty}^{0} dt_1 \int_{-\infty}^{0} dt_2 A_1(x_1,t_1)A_1(x_2,t_2)\langle 0|\left[\partial\hat{J}_1(x_1,t_1)/\partial t_1, \hat{J}_1(x_2,t_2)\right]|0\rangle \tag{6.7}$$

This can be re-written as,

$$\xi_f = \frac{i}{2}\int dx_1\int dx_2 \int_{-\infty}^{0} dt_1 \int_{-\infty}^{0} dt_2 A_1(x_1,t_1)A_1(x_2,t_2)\frac{\partial}{\partial t_1}\langle 0|\left[\hat{J}_1(x_1,t_1), \hat{J}_1(x_2,t_2)\right]|0\rangle \tag{6.8}$$

Use $\left(\theta(t_1-t_2)+\theta(t_2-t_1)\right)=1$ to obtain,

$$\xi_f = \frac{i}{2}\int dx_1\int dx_2 \int_{-\infty}^{0} dt_1 \int_{-\infty}^{0} dt_2 \left[A_1(x_1,t_1)A_1(x_2,t_2)\frac{\partial}{\partial t_1}\left(\langle 0|\left[\hat{J}_1(x_1,t_1), \hat{J}_1(x_2,t_2)\right]|0\rangle \begin{pmatrix}\theta(t_1-t_2) \\ +\theta(t_2-t_1)\end{pmatrix}\right)\right] \tag{6.9}$$

Use Eq. (5.5) in the above to obtain,

$$\xi_f = -\frac{1}{2}\int dx_1\int dx_2 \int_{-\infty}^{0} dt_1 \int_{-\infty}^{0} dt_2 \left[A_1(x_1,t_1)A_1(x_2,t_2)\frac{\partial}{\partial t_1}\begin{pmatrix}\pi_{11}(x_1-x_2,t_1-t_2) \\ -\pi_{11}(x_2-x_1,t_2-t_1)\end{pmatrix}\right] \tag{6.10}$$

Next use $\partial\pi_{11}(x_2-x_1,t_2-t_1)/\partial t_1 = -\partial\pi_{11}(x_2-x_1,t_2-t_1)/\partial t_2$ to obtain,

$$\xi_f = -\frac{1}{2}\int dx_1 \int dx_2 \int_{-\infty}^{0} dt_1 \int_{-\infty}^{0} dt_2 \left[ A_1(x_1,t_1) A_1(x_2,t_2) \left( \frac{\partial \pi_{11}(x_1 - x_2, t_1 - t_2)}{\partial t_1} + \frac{\partial \pi_{11}(x_2 - x_1, t_2 - t_1)}{\partial t_2} \right) \right] \tag{6.11}$$

Rearrange the dummy integration variables to obtain,

$$\xi_f = -\int dx_1 \int dx_2 \int_{-\infty}^{0} dt_1 A_1(x_1,t_1) \frac{d}{dt_1} \int_{-\infty}^{+\infty} dt_2 \left[ A_1(x_2,t_2) \pi_{11}(x_1 - x_2, t_1 - t_2) \right] \tag{6.12}$$

Note that the upper limit of integration with respect to $t_2$ has been changed from $0$ to $\infty$. This is allowed because the vacuum polarization tensor contains a $\theta(t_1 - t_2)$ term and the upper limit on $t_1$ is $0$.

So far we have just rearranged terms. If we use Eq. (5.7) and (5.8) for $\pi_{11}$ in the above we should expect to obtain (6.4). However recall that to obtain a gauge invariant theory we had to replace $\pi_{\mu\nu}(k_1,k_0)$ with $\bar{\pi}_{\mu\nu}(k_1,k_0)$. Therefore, to be mathematically consistent, we should replace $\pi_{11}(x,t)$, in (6.12), with $\bar{\pi}_{11}(x,t)$ to obtain,

$$\bar{\xi}_f = -\int dx_1 \int dx_2 \int_{-\infty}^{0} dt_1 A_1(x_1,t_1) \frac{\partial}{\partial t_1} \int_{-\infty}^{+\infty} dt_2 \left[ A_1(x_2,t_2) \bar{\pi}_{11}(x_1 - x_2, t_1 - t_2) \right] \tag{6.13}$$

where $\bar{\xi}_f$ is now considered to be the "regularized" free field energy. From (5.14) we have,

$$\bar{\pi}_{11}(x,t) = \pi_{11}(x,t) - \Gamma \delta(x) \delta(t) \tag{6.14}$$

Using this $\bar{\xi}_f$ can be written as,

$$\bar{\xi}_f = \xi_f - \Delta\xi_f \tag{6.15}$$

where,

$$\Delta\xi_f = -\int dx_1 \int dx_2 \int_{-\infty}^{0} dt_1 A_1(x_1,t_1) \frac{\partial}{\partial t_1} \int_{-\infty}^{+\infty} dt_2 \left[ A_1(x_2,t_2) \Gamma \delta(x_1 - x_2) \delta(t_1 - t_2) \right] \tag{6.16}$$

This is evaluated as,

$$\Delta\xi_f = -\Gamma \int dx_1 \int_{-\infty}^{0} dt_1 A_1(x_1,t_1) \frac{\partial A_1(x_1,t_1)}{\partial t_1} = \frac{1}{2} |\Gamma| \int dx_1 \left( A_1(x_1,0) \right)^2 \tag{6.17}$$

where, due to the fact that $\Gamma$ is negative, we have replaced $-\Gamma$ with $|\Gamma|$ to emphasis that $\Delta\xi_f$ is positive. The result is that the regularized free field energy $\bar{\xi}_f$ is reduced by the amount $\Delta\xi_f$ from the original value $\xi_f$.

Next recall that we have already solved for $\xi_f$ for the case where $A_0 = 0$ and $A_1 = fe^{\lambda t}/\sqrt{L}$ (see Eq. (6.4)). However we have found a new expression for $\xi_f$ in terms of $\pi_{11}(x,t)$ (see Eq. (6.12)). From Eq. (8.14) of Appendix 1, $\pi_{11}(x,t)$ is,

$$\pi_{11}(x,t) = -i\theta(t)\sum_{p,q} F_{11}(q,p)\left[e^{+i(p+q)x}e^{-i(E_p+E_q)t} - e^{-i(p+q)x}e^{+i(E_p+E_q)t}\right] \tag{6.18}$$

When this is used in (6.12) along with $A_1 = fe^{\lambda t}/\sqrt{L}$ we obtain the same result for $\xi_f$ as Eq. (6.4) which confirms the validity of the new expression.

Next, use $A_1(x,0) = f/\sqrt{L}$ in Eq. (6.17) to obtain $\Delta\xi_f = f^2\frac{1}{2}|\Gamma| = f^2e^2/(2\pi)$ since $|\Gamma| = (e^2/\pi)$ (see Eq. (5.11)). Therefore,

$$\bar{\xi}_f = \xi_f - \Delta\xi_f = \left\{\frac{f^2e^2m^2}{4\pi}\int_{-\infty}^{+\infty}\left(\frac{1}{E_p\left(E_p^{\,2} + (\lambda/2)^2\right)}\right)dp\right\} - \frac{f^2e^2}{2\pi} \tag{6.19}$$

Next, it can be shown that $1 = (m^2/2)\int_{-\infty}^{+\infty}(dp/E_p^{\,3})$. Use this in the above to obtain,

$$\bar{\xi}_f = \xi_f - \Delta\xi_f = \frac{f^2e^2m^2}{4\pi}\int_{-\infty}^{+\infty}\left(\frac{1}{E_p\left(E_p^{\,2} + (\lambda/2)^2\right)} - \frac{1}{E_p^{\,3}}\right)dp < 0 \tag{6.20}$$

This is negative because the integrand is always negative if $\lambda \neq 0$. Therefore the use of regularization to produce a gauge invariant theory also results in a theory where there exists quantum states whose free field energy is less than that of the vacuum state.

## 7. Pauli-Villars regularization.

In this section we will apply the Pauli-Villars regularization method in order to obtain a gauge invariant vacuum polarization tensor. To apply the Pauli-Villars method for our 1+1 dimensional problem we first assume that there exists a "fictitious" fermion field with infinite mass. Next we calculate the polarization tensor for this fermion field which will be $\pi_{\mu\nu}(k_1,k_0;m\to\infty)$. The quantity $\pi_{\mu\nu}(k_1,k_0;m\to\infty)$ is simply $\pi_{\mu\nu}(k_1,k_0)$ as calculated in Section 5 (see Eq. (5.8)) with $m\to\infty$. Next subtract $\pi_{\mu\nu}(k_1,k_0;m\to\infty)$ from $\pi_{\mu\nu}(k_1,k_0)$ to obtain the regularization vacuum polarization tensor, i.e.,

$$\bar{\pi}_{PV,\mu\nu}(k_1,k_0) = \pi_{\mu\nu}(k_1,k_0) - \pi_{\mu\nu}(k_1,k_0;m\to\infty) \tag{7.1}$$

It can be shown that $\pi_{\mu\nu}\left(k_1,k_0;m\to\infty\right)=\pi_{\mu\nu,NG}\left(k_1,k_0\right)$. Therefore $\bar{\pi}_{PV,\mu\nu}\left(k_1,k_0\right)=\bar{\pi}_{\mu\nu}\left(k_1,k_0\right)$ which is the original regularized vacuum polarization tensor as discussed at the end of Section 5.

So far Pauli-Villars regularization hasn't changed anything. The advantage of Pauli-Villars regularization is that it is useful in calculating other quantities. Consider the calculation of the free field energy given by Eq. (6.3). In Section 6 we had to go through some effort to obtain (6.12) in order to express the free field energy as a function of $\pi_{\mu\nu}$. However if we use Pauli-Villars regularization to regularize the vacuum polarization tensor then for mathematical consistency we need to use it for the regularizing the energy. In this case we obtain $\bar{\xi}_{f,PV}=\xi_f-\xi_f\left(m\to\infty\right)$ were $\xi_f$ was given by Eq. (6.4). In Appendix 3 it is shown that,

$$\xi_f\left(m\to\infty\right)=\left(\frac{f^2e^2}{2\pi}\right) \tag{7.2}$$

Therefore $\xi_f\left(m\to\infty\right)=\Delta\xi_f$ (see discussion at the end of Section 6) and $\bar{\xi}_{f,PV}=\bar{\xi}_f$ (Eq. (6.20) which is the result obtained at the end of the last section. Therefore the use of Pauli-Villars regularization leads to the previously obtained result that there exist quantum state with less free field energy than the vacuum state.

## 8. Conclusion.

In this paper we have investigated the fact that a direct calculation of the vacuum polarization tensor will include a non-gauge invariant part. This is a problem because the theory is supposed to be gauge invariant. It has been suggested that this problem may be due to an "artifact of the mathematics". However it has been shown that this is not the case. This problem arises because there is a mathematical incompatibility between the requirement of gauge invariance and the fact that the theory is formulated in such a way that there is a lower bound to the free field energy. Essentially the problem is built into the theory at a fundamental level. To illustrate this fact the vacuum polarization tensor is calculated in 1+1 dimensional space-time and was shown to contain a non-gauge invariant part. In order to achieve a gauge invariant theory this non-gauge invariant part must be removed. This will make the theory gauge invariant.

It was then shown in Section 6 that this process of "regularizing" the vacuum polarization tensor requires that the free field energy be regularized also. This will result in the existence of state vectors which have less energy than the vacuum state $|0\rangle$. This is consistent with our discussion in Section 3 where it was shown that there is an incompatibility between the requirement that the theory be gauge invariant and that the free field energy has a lower bound.

## Appendix 1.

We will evaluate the vacuum polarization tensor which is given in equation (5.5) and reproduced below,

$$\pi_{\mu\nu}\left(x-x_1,t-t_1\right)=-i\left\langle 0\left|\left[\hat{J}_{\mu}\left(x,t\right),\hat{J}_{\nu}\left(x_1,t_1\right)\right]\right|0\right\rangle\theta\left(t-t_1\right) \tag{8.1}$$

The field operator in the interaction picture is given by,

$$\hat{\psi}\left(x,t\right)=\sum_{p}\left(\hat{b}_p u_p e^{ipx}e^{-iE_pt}+\hat{d}_p^{\dagger}v_p e^{-ipx}e^{+iE_pt}\right),\quad \hat{\psi}^{\dagger}\left(x,t\right)=\sum_{p}\left(\hat{b}_p^{\dagger}u_p^{\dagger}e^{-ipx}e^{+iE_pt}+\hat{d}_p v_p^{\dagger}e^{+ipx}e^{-iE_pt}\right) \tag{8.2}$$

The interaction picture current operators are given by,

$$\hat{J}_1\left(x,t\right)=e:\psi^{\dagger}\left(x,t\right)\sigma_1\psi\left(x,t\right):,\quad \hat{J}_0\left(x,t\right)=e:\psi^{\dagger}\left(x,t\right)\psi\left(x,t\right): \tag{8.3}$$

Using the above and the information in Section 4 $\pi_{\mu\nu}\left(x-x_1,t-t_1\right)$ can be evaluated as follows. First we can show that,

$$\hat{J}_{\nu}\left(x_1,t_1\right)\left|0\right\rangle=e\sum_{p,q}\left(u_p^{\dagger}\alpha_{\nu}v_q\right)e^{-i(p+q)x_1}e^{+i\left(E_p+E_q\right)t_1}\hat{b}_p^{\dagger}\hat{d}_q^{\dagger}\left|0\right\rangle \tag{8.4}$$

$$\left\langle 0\right|\hat{J}_{\beta}\left(x,t\right)=e\left\langle 0\right|\hat{d}_q\hat{b}_p\sum_{p,q}\left(v_q^{\dagger}\alpha_{\beta}u_p\right)e^{+i(p+q)x}e^{-i\left(E_p+E_q\right)t} \tag{8.5}$$

where $\alpha_{\nu}$ are the $2\times 2$ matrices, $\alpha_1=\sigma_1$ and $\alpha_0=I$. From this we obtain,

$$\left\langle 0\right|\hat{J}_{\beta}\left(x,t\right)\hat{J}_{\nu}\left(x_1,t_1\right)\left|0\right\rangle=e^2\sum_{p,q}F_{\beta\nu}\left(q,p\right)e^{+i(p+q)(x-x_1)}e^{-i\left(E_p+E_q\right)(t-t_1)} \tag{8.6}$$

where $F_{\beta\nu}\left(q,p\right)=\left(v_q^{\dagger}\alpha_{\beta}u_p\right)\left(u_p^{\dagger}\alpha_{\nu}v_q\right)$. Note that,

$$\left\langle 0\left|\left[\hat{J}_{\beta}\left(x,t\right),\hat{J}_{\nu}\left(x_1,t_1\right)\right]\right|0\right\rangle=\left\langle 0\right|\hat{J}_{\beta}\left(x,t\right)\hat{J}_{\nu}\left(x_1,t_1\right)\left|0\right\rangle-(c.c.) \tag{8.7}$$

Use (4.4) to obtain the following,

$$F_{00}\left(q,p\right)=N_q^{\ 2}N_p^{\ 2}\left(\frac{q}{E_q+m}+\frac{p}{E_p+m}\right)^2 \tag{8.8}$$

$$F_{10}\left(q,p\right)=F_{01}\left(q,p\right)=N_q^{\ 2}N_p^{\ 2}\left(\frac{q}{E_q+m}+\frac{p}{E_p+m}\right)\left(1+\frac{pq}{\left(E_p+m\right)\left(E_q+m\right)}\right) \tag{8.9}$$

$$F_{11}\left(q,p\right)=N_q^{\ 2}N_p^{\ 2}\left(1+\frac{pq}{\left(E_p+m\right)\left(E_q+m\right)}\right)^2 \tag{8.10}$$

This can be further evaluated to obtain,

$$F_{00}(q,p)=\frac{1}{2L^2E_pE_q}\left(qp+E_qE_p-m^2\right) \tag{8.11}$$

$$F_{10}(q,p)=F_{01}(q,p)=\frac{qE_p+pE_q}{2L^2E_pE_q} \tag{8.12}$$

$$F_{11}(q,p)=\frac{1}{2L^2E_pE_q}\left(qp+E_qE_p+m^2\right) \tag{8.13}$$

Use the above results in (8.1) to obtain,

$$\pi_{\beta\nu}(x-x_1,t-t_1)=-i\theta(t-t_1)e^2\sum_{p,q}F_{\beta\nu}(q,p)\begin{bmatrix} e^{+i(p+q)(x-x_1)}e^{-i(E_p+E_q)(t-t_1)} \\ -e^{-i(p+q)(x-x_1)}e^{+i(E_p+E_q)(t-t_1)}\end{bmatrix} \tag{8.14}$$

Use the following expression for the Fourier transform,

$$\pi_{\beta\nu}(k_1,k_0)=\int\pi_{\beta\nu}(x,t)e^{-ik_0t}e^{+ik_1x}dxdt \tag{8.15}$$

and (8.14) to obtain,

$$\pi_{\beta\nu}(k_1,k_0)=-ie^2\sum_{pq}F_{\beta\nu}(q,p)\int dxe^{+ik_1x}\int_0^\infty dt\begin{bmatrix} e^{+i(p+q)x}e^{-i(E_p+E_q)t} \\ -e^{-i(p+q)x}e^{+i(E_p+E_q)t}\end{bmatrix}e^{-ik_0t}e^{-\varepsilon t} \tag{8.16}$$

where $\varepsilon\to 0$ from above and is there to ensure the integral is well-defined at the limit $t=\infty$. Perform the integration to obtain,

$$\pi_{\beta\nu}(k_1,k_0)=-ie^2\sum_{pq}F_{\beta\nu}(q,p)\left(\frac{-L\delta_{k_1+p+q}}{-i\left[\left(E_p+E_q\right)+k_0\right]-\varepsilon}+\frac{L\delta_{k_1-p-q}}{i\left[\left(E_p+E_q\right)-k_0\right]-\varepsilon}\right) \tag{8.17}$$

Note that the factor of $L$ is due to the fact that integration with respect to $x$ is over the integration volume $L\to\infty$. This yields,

$$\pi_{\beta\nu}(k_1,k_0)=-e^2L\sum_p\left(\frac{F_{\beta\nu}\left(-(p+k_1),p\right)}{\left[\left(E_p+E_{p+k_1}\right)+k_0\right]-i\varepsilon}+\frac{F_{\beta\nu}\left((k_1-p),p\right)}{\left[\left(E_p+E_{p-k_1}\right)-k_0\right]+i\varepsilon}\right) \tag{8.18}$$

Next make the substitution $p\to -p$ in the second term on the right to obtain,

$$\pi_{\beta\nu}(k_1,k_0)=-e^2L\sum_p\left(\frac{F_{\beta\nu}\left(-(p+k_1),p\right)}{\left[\left(E_p+E_{p+k_1}\right)+k_0\right]-i\varepsilon}+\frac{F_{\beta\nu}\left((p+k_1),-p\right)}{\left[\left(E_p+E_{p+k_1}\right)-k_0\right]+i\varepsilon}\right) \tag{8.19}$$

Note that $F_{00}(-q,p)=F_{00}(q,-p)$, $F_{11}(-q,p)=F_{11}(q,-p)$, $F_{10}(-q,p)=-F_{10}(q,-p)$. This yields,

$$\pi_{00}\left(k_1,k_0\right)=-e^2L\sum_p F_{00}\left(-\left(p+k_1\right),p\right)\left(\frac{2\left(E_p+E_{p+k_1}\right)}{\left[\left(E_p+E_{p+k_1}\right)^2-k_0{}^2\right]+2k_0 i\varepsilon}\right) \tag{8.20}$$

$$\pi_{10}\left(k_1,k_0\right)=-e^2L\sum_p F_{10}\left(-\left(p+k_1\right),p\right)\left(\frac{-2k_0}{\left[\left(E_p+E_{p+k_1}\right)^2-k_0{}^2\right]+2k_0 i\varepsilon}\right) \tag{8.21}$$

$$\pi_{11}\left(k_1,k_0\right)=-e^2L\sum_p F_{11}\left(-\left(p+k_1\right),p\right)\left(\frac{2\left(E_p+E_{p+k_1}\right)}{\left[\left(E_p+E_{p+k_1}\right)^2-k_0{}^2\right]+2k_0 i\varepsilon}\right) \tag{8.22}$$

Convert the summation to an integral by using $\sum_p \to \int \frac{L}{2\pi}dp$ and make the substitution $p \to p - k_1/2$ to obtain,

$$\pi_{00}\left(k_1,k_0\right)=-\frac{e^2}{2\pi}\int dp\frac{\left(E_{p-k_1/2}E_{p+k_1/2}-\left(p-k_1/2\right)\left(p+k_1/2\right)-m^2\right)}{2E_{p-k_1/2}E_{p+k_1/2}}\left(\frac{2\left(E_{p+k_1/2}+E_{p-k_1/2}\right)}{D\left(p,k_1,k_0\right)}\right) \tag{8.23}$$

$$\pi_{10}\left(k_1,k_0\right)=-\frac{e^2}{2\pi}\int dp\frac{\left(-\left(p+k_1/2\right)E_{p-k_1/2}+\left(p-k_1/2\right)E_{p+k_1/2}\right)}{2E_{p-k_1/2}E_{p+k_1/2}}\left(\frac{-2k_0}{D\left(p,k_1,k_0\right)}\right) \tag{8.24}$$

$$\pi_{11}\left(k_1,k_0\right)=-\frac{e^2}{2\pi}\int dp\frac{\left(E_{p-k_1/2}E_{p+k_1/2}-\left(p-k_1/2\right)\left(p+k_1/2\right)+m^2\right)}{2E_{p-k_1/2}E_{p+k_1/2}}\left(\frac{2\left(E_{p+k_1/2}+E_{p-k_1/2}\right)}{D\left(p,k_1,k_0\right)}\right) \tag{8.25}$$

where $D\left(p,k_1,k_0\right)=\left[\left(E_{p+k_1/2}+E_{p-k_1/2}\right)^2-k_0{}^2\right]+2k_0 i\varepsilon$. Next make the substitutions,

$$u=\frac{1}{2}\left(E_{p+k_1/2}+E_{p-k_1/2}\right),\quad v=\frac{1}{2}\left(E_{p+k_1/2}-E_{p-k_1/2}\right) \tag{8.26}$$

From this we obtain,

$$v^2=\frac{k_1{}^2}{4}\left[1-\frac{m^2}{\left(u^2-k_1{}^2/4\right)}\right] \tag{8.27}$$

and,

$$p=2uv/k_1 \tag{8.28}$$

$$du=\frac{1}{2}\left(\frac{p+k_1/2}{E_{p+k_1/2}}+\frac{p-k_1/2}{E_{p-k_1/2}}\right)dp=\frac{\left(2up-k_1v\right)}{2E_{p+k_1/2}E_{p-k_1/2}}dp \tag{8.29}$$

This yields,

$$dp = \frac{2E_{p+k_1/2}E_{p-k_1/2}}{\left(2up - k_1 v\right)} du \tag{8.30}$$

Substitute all this into (8.23) to obtain,

$$\pi_{00}\left(k_1,k_0\right) = -\frac{e^2 m^2}{4\pi} k_1^{\ 2} \int\limits_{\sqrt{m^2+k_1^{\ 2}/4}}^{\infty} \frac{udu}{\left(u^2-\left(k_1^{\ 2}/4\right)\right)^2 \sqrt{1-\left[m^2/\left(u^2-k_1^{\ 2}/4\right)\right]}} \left(\frac{1}{u^2-\left(k_0^{\ 2}/4\right)+\left(k_0 i\varepsilon/2\right)}\right) \tag{8.31}$$

Define $s = u^2 - \left(k_1^{\ 2}/4\right)$ which yields $u^2 = s + \left(k_1^{\ 2}/4\right)$. Use this in the above to obtain,

$$\pi_{00}\left(k_1,k_0\right) = k_1^{\ 2}\Pi\left(k^2\right) \tag{8.32}$$

where,

$$\Pi\left(k^2\right) = -\frac{e^2 m^2}{8\pi} \int\limits_{m^2}^{\infty} ds \left(\frac{1}{s^2\sqrt{1-\left(m^2/s\right)}}\right)\left(\frac{1}{s-\left(k^2/4\right)+\left(k_0 i\varepsilon/2\right)}\right) \tag{8.33}$$

Recall $k^2 = k_0^{\ 2} - k_1^{\ 2}$. Similarly we obtain,

$$\pi_{10}\left(k_1,k_0\right) = \pi_{01}\left(k_1,k_0\right) = k_1 k_0 \Pi\left(k^2\right) \tag{8.34}$$

and,

$$\pi_{11}\left(k_1,k_0\right) = k_0^{\ 2}\Pi\left(k^2\right) + \Gamma \tag{8.35}$$

where,

$$\Gamma = \frac{-e^2 m^2}{2\pi} \int\limits_{m^2}^{\infty} ds \left(\frac{1}{s^2\sqrt{1-\left(m^2/s\right)}}\right) = -\frac{e^2}{\pi} \tag{8.36}$$

The above relationships can be rewritten as,

$$\pi_{\mu\nu}\left(k_1,k_0\right) = \pi_{\mu\nu,G}\left(k_1,k_0\right) + \pi_{\mu\nu,NG}\left(k_1,k_0\right) \tag{8.37}$$

where,

$$\pi_{\mu\nu,G}\left(k_1,k_0\right) = \left(k_\mu k_\nu - g_{\mu\nu}k^2\right)\Pi\left(k^2\right), \quad \pi_{\mu\nu,NG}\left(k_1,k_0\right) = g_\mu^1 g_\nu^1 \Gamma \tag{8.38}$$

where $g_\rho^1 = 1$ if $\rho = 1$ and $g_\rho^1 = 0$ if $\rho = 0$.

Note that $k^u \pi_{\mu\nu,G}\left(k_1,k_0\right) = 0$ and $k^u \pi_{\mu\nu,NG}\left(k_1,k_0\right) = k^1 g_\nu^1 \Gamma \neq 0$ if $\nu = 1$. Therefore we have found that the vacuum polarization tensor $\pi_{\mu\nu}\left(k_1,k_0\right)$ is not gauge invariant. It consists of two terms where $\pi_{\mu\nu,G}\left(k_1,k_0\right)$ is the gauge invariant part and $\pi_{\mu\nu,NG}\left(k_1,k_0\right)$ is the non-gauge invariant part.

Next we will rewrite $\Pi\left(k^2\right)$ in a form that is more consistent with some of the literature. Let $s=m^2/\left[4y(1-y)\right]$. From this we obtain $\sqrt{1-\left(m^2/s\right)}=(1-2y)$ along with $ds=-\left\{\left(m^2/4\right)(1-2y)/\left[y(1-y)\right]^2\right\}dy$. Use this in (8.33) to obtain,

$$\Pi\left(k^2\right)=-\frac{2e^2}{\pi}\int_0^{1/2}\left(\frac{y(1-y)dy}{m^2-k^2y(1-y)+k_0 i\varepsilon 2y(1-y)}\right)=-\frac{e^2}{\pi}\int_0^1\left(\frac{y(1-y)dy}{m^2-k^2y(1-y)+k_0 i\varepsilon}\right) \tag{8.39}$$

The last relationship is due to the fact that $y(1-y)$ is symmetric around $y=1/2$ and we replace $\varepsilon 2y(1-y)$ with $\varepsilon$.

## Appendix 2.

We want to evaluate Eq. (6.3) which, for convenience, is reproduced below,

$$\xi_f=\left(f^2/L\right)\langle 0|\left(\int_{-\infty}^{0}e^{\lambda t}dt\int dx\hat{J}_1(x,t)\right)\hat{H}_0\left(\int_{-\infty}^{0}e^{\lambda t}dt\int dx\hat{J}_1(x,t)\right)|0\rangle \tag{9.1}$$

Using the results of Section 4 we can show,

$$\left(\int_{-\infty}^{0}dt\int dxA_1(x,t)\hat{J}_1(x,t)\right)|0\rangle=\left(e\int_{-\infty}^{0}dt\int dx\sum_{p,q}\left(u_p^\dagger\alpha_1 v_q\right)\hat{b}_p^\dagger\hat{d}_q^\dagger e^{-i(p+q)x}e^{i\left(E_p+E_q\right)t}A_1(x,t)\right)|0\rangle \tag{9.2}$$

Let $A_1(x,t)=fe^{\lambda t}/\sqrt{L}$. Therefore,

$$\left(f/\sqrt{L}\right)\left(\int_{-\infty}^{0}e^{\lambda t}dt\int dx\hat{J}_1(x,t)\right)|0\rangle=ef\sqrt{L}\left(\int_{-\infty}^{0}dt\sum_p\left(u_p^\dagger\alpha_1 v_{-p}\right)\hat{b}_p^\dagger\hat{d}_{-p}^\dagger e^{i2E_pt}e^{\lambda t}\right)|0\rangle \tag{9.3}$$

To obtain the above result we use $\int dxe^{-i(p+q)x}=\delta_{p+q}L$. From Section 4 we obtain,

$$u_p^\dagger\alpha_1 v_{-p}=N_pN_{-p}\left(1-\frac{p^2}{\left(E_p{}^2+m^2\right)}\right)=\frac{m}{E_pL} \tag{9.4}$$

Use this in (9.3) to obtain,

$$\left(f/\sqrt{L}\right)\left(\int_{-\infty}^{0}e^{\lambda t}dt\int dx\hat{J}_1(x,t)\right)|0\rangle=\left(ef/\sqrt{L}\right)\sum_p\frac{m}{\left(\lambda+i2E_p\right)E_p}\hat{b}_p^\dagger\hat{d}_{-p}^\dagger|0\rangle \tag{9.5}$$

Also we can show that,

$$\left(f/\sqrt{L}\right)\langle 0|\left(\int_{-\infty}^{0}e^{\lambda t}dt\int dx\hat{J}_1(x,t)\right)=\left(ef/\sqrt{L}\right)\langle 0|\sum_p\hat{d}_{-p}\hat{b}_p\frac{m}{\left(\lambda-i2E_p\right)E_p} \tag{9.6}$$

Use all this along with $\langle 0|\hat{d}_{-p}\hat{b}_p\hat{H}_0\hat{b}_p^\dagger\hat{d}_{-p}^\dagger|0\rangle = 2E_p$ in (9.1) to obtain,

$$\xi_f = f^2\frac{e^2m^2}{2L}\sum_p\frac{1}{\left(E_p{}^2+\left(\lambda^2/4\right)\right)E_p} \tag{9.7}$$

This can be expressed in integral form to obtain,

$$\xi_f = f^2\frac{e^2m^2}{2}\left(\frac{1}{2\pi}\right)\int_{-\infty}^{+\infty}\frac{dp}{\left(E_p{}^2+\left(\lambda^2/4\right)\right)E_p} \tag{9.8}$$

## Appendix 3.

We want to evaluate (6.4) in the limit $m\to\infty$. Perform the integration of (6.4) to obtain,

$$\xi_f = \left|\left(\frac{f^2}{2\pi}\right)\frac{e^2m^2}{2}\left(\ln\left\{\frac{\left|\left[(\lambda/2)mp\big/\sqrt{m^2+p^2}\right]+m\sqrt{m^2+(\lambda/2)^2}\right|}{\left|\left[(\lambda/2)mp\big/\sqrt{m^2+p^2}\right]-m\sqrt{m^2+(\lambda/2)^2}\right|}\right\}\right)\Bigg/\left(\lambda\sqrt{m^2+(\lambda/2)^2}\right)\right|_{-\infty}^{+\infty} \tag{10.1}$$

This becomes,

$$\xi_f = \frac{e^2m^2}{\lambda\sqrt{m^2+(\lambda/2)^2}}\left(\frac{f^2}{2\pi}\right)\ln\left\{\frac{m\sqrt{m^2+(\lambda/2)^2}+(\lambda m/2)}{m\sqrt{m^2+(\lambda/2)^2}-(\lambda m/2)}\right\} \tag{10.2}$$

To determine $\xi_f(m\to\infty)$ use $\sqrt{m^2+(\lambda/2)^2} \underset{m\to\infty}{=} m\left(1+\lambda^2/8m^2\right)$ in the above to obtain,

$$\xi_f(m\to\infty) \underset{m\to\infty}{=} \frac{e^2m}{\lambda\left(1+\lambda^2/8m^2\right)}\left(\frac{f^2}{2\pi}\right)\ln\left\{\frac{m^2\left(1+\lambda^2/8m^2\right)+(\lambda m/2)}{m^2\left(1+\lambda^2/8m^2\right)-(\lambda m/2)}\right\} \tag{10.3}$$

Take $m\to\infty$ to obtain,

$$\xi_f(m\to\infty) \underset{m\to\infty}{=} \frac{e^2m}{\lambda}\left(\frac{f^2}{2\pi}\right)\ln\left\{\frac{1+(\lambda/2m)}{1-(\lambda/2m)}\right\} \underset{m\to\infty}{=} \frac{m}{\lambda}\left(\frac{f^2}{2\pi}\right)\left(\frac{\lambda}{m}\right)=\left(\frac{f^2}{2\pi}\right) \tag{10.4}$$

## Appendix 4.

Next we will examine $\pi_{T,\mu\nu}$ which has been defined as,

$$\pi_{T,\mu\nu}(x-x_1,t-t_1) = -i\langle 0|T\left[\hat{J}_\mu(x,t)\hat{J}_\nu(x_1,t_1)\right]|0\rangle \tag{11.1}$$

The evaluation of this quantity is similar to the evaluation of $\pi_{\mu\nu}$ which was done in Appendix 1. It can be shown that,

$$\pi_{T,\mu\nu}(k_1,k_0) = \pi_{T,\mu\nu,G}(k_1,k_0)+\pi_{T,\mu\nu,NG}(k_1,k_0) \tag{11.2}$$

where,

$$\pi_{T,\mu\nu,G}\left(k_1,k_0\right)=\left(k_\mu k_\nu-g_{\mu\nu}k^2\right)\Pi_T\left(k^2\right),\quad \pi_{T,\mu\nu,NG}\left(k_1,k_0\right)=\pi_{\mu\nu,NG}\left(k_1,k_0\right)=g_\mu^1 g_\nu^1\Gamma \tag{11.3}$$

and where,

$$\Pi_T\left(k^2\right)=-\frac{e^2m^2}{8\pi}\int_{m^2}^{\infty}ds\left(\frac{1}{s^2\sqrt{1-\left(m^2/s\right)}}\right)\left(\frac{1}{s-\left(k^2/4\right)+i\varepsilon}\right) \tag{11.4}$$

This can also be rewritten as,

$$\Pi_T\left(k^2\right)=-\frac{e^2}{\pi}\int_0^1\left(\frac{y(1-y)dy}{m^2-k^2y(1-y)+i\varepsilon}\right) \tag{11.5}$$

In order to obtain a gauge invariant quantity we define the regularized quantity $\bar{\pi}_{T,\mu\nu}$,

$$\bar{\pi}_{T,\mu\nu}\left(k_1,k_0\right)=\pi_{T,\mu\nu}\left(k_1,k_0\right)-\pi_{T,\mu\nu,NG}\left(k_1,k_0\right)=\pi_{T,\mu\nu,G}\left(k_1,k_0\right) \tag{11.6}$$

## References.

[1] R.A. Bertlmann. *Anomalies in Quantum Field Theory.* 2000 (Oxford University Press Inc., New York).

[2] S.B. Treiman, R. Jackiw, B. Zumino, E. Witten. *Current Algebra and Anomalies*. 1985 (Princeton University Press, Princeton University Press).

[3] J.A. Harvey. "TASI 2003 Lectures on Anomalies". arXiv:hep-th/0509097.

[4] D. Solomon. "A new look at the problem of gauge invariance in quantum field theory". *Phys. Scr*. **76** 64 (2007).

[5] D. Solomon. "Gauge invariance and the vacuum state". *Can. J. Phys.* **76** 111 (1998).

[6] C. Jarlskog. "Supersymetry – Early Roots That Didn't Grow". Advances in High Energy Physics. Article ID 764875. May 2015. Also – arxiv: 1503.07629v1.

[7] J. Schwinger. "On gauge invariance and vacuum polarization". Phys. Rev. **82** 664 (1951).

[8] W. Greiner, B. Muller, and T. Rajelski. *Quantum Electrodynamics of Strong Fields.* 1985. (Springer, Berlin).

[9] W. Heitler. *The quantum theory of radiation.* 1954 (Dover Publications, New York).

[10] W. Pauli. *Pauli Lectures on Physic: Volume 6. Selected Topics in Field Quantization.* (1976) MIT Press, Cambridge, Massachusetts and London, England).

[11] K. Nishijima. *Fields and Particles: Field theory and Dispersion Relations*. 1969 (W.A. Benjamin, New York).

[12] J.J. Sakurai. *Advanced Quantum Mechanics.* 1967 (Addison-Wesley Publishing Co., Redwood City, California).

[13] R. Jost and J. Rayski, "Remark on the problem of the vacuum polarization and the photon self-energy". Helv. Phys. Acta. (1949), 457.

[14] W. Greiner and J. Reinhardt. *Quantum Electrodynamics*. 1992 (Berlin: Springer).

[15] M.E. Peskin and D.V. Schroeder. *An Introduction to quantum field theory.* 1995 (Reading, MA: Addison-Wesley).